\documentclass[sigconf]{acmart} 

\acmConference[ASE 2022]{IEEE/ACM Automated Software Engineering}{October 10-14, 2022}{Michigan, United States}
\usepackage{color}
\usepackage{etoolbox}
\usepackage{listings}
\usepackage[T1]{fontenc}

\usepackage[scaled=0.8]{beramono}

\newcommand{\lstfontfamily}{\ttfamily}

\definecolor{darkviolet}{rgb}{0.5,0,0.4}
\definecolor{darkgreen}{rgb}{0,0.4,0.2} 
\definecolor{darkblue}{rgb}{0.1,0.1,0.9}
\definecolor{darkgrey}{rgb}{0.5,0.5,0.5}
\definecolor{lightblue}{rgb}{0.4,0.4,1}

\definecolor{stringColor}{rgb}{0.16,0.00,1.00}
\definecolor{annotationColor}{rgb}{0.39,0.39,0.39}
\definecolor{keywordColor}{rgb}{0.50,0.00,0.33}
\definecolor{commentColor}{rgb}{0.25,0.50,0.37}
\definecolor{javadocColor}{rgb}{0.25,0.37,0.75}
\definecolor{jTagColor}{rgb}{0.50,0.62,0.75}
\definecolor{eTagColor}{rgb}{0.50,0.62,0.75}
\definecolor{lineNumberColor}{rgb}{0.47,0.47,0.47}

\def\jTags{@author, @deprecated, @exception, @param, @return, @see, @serial, @serialData, @serialField, @since, @throws, @version}

\def\jAnnotations{
    classoffset=1,
    morekeywords={@Override, @Deperecated, @SuppressWarnings, @Retention, @Documented, @Target, @Inherited},
    keywordstyle=\color{annotationColor},
    classoffset=0
}

\def\eTags{FIXME, TODO, XXX}

\newrobustcmd{\markupJavadocs}[1]{%
\edef\mytok{\the\lst@token}%
\renewcommand*{\do}[1]{%
\ifdefstring{\mytok}{##1}%
{\color{jTagColor}\bfseries\listbreak}%
{}%
}%
{\color{javadocColor}%
\expandafter\docsvlist\expandafter{\jTags}%
\renewcommand*{\do}[1]{%
\ifdefstring{\mytok}{##1}%
{\color{eTagColor}\bfseries\listbreak}%
{}%
}%
\expandafter\docsvlist\expandafter{\eTags}%
#1}%
}%

\newrobustcmd{\markupComments}[1]{%
\edef\mytok{\the\lst@token}%
\renewcommand*{\do}[1]{%
\ifdefstring{\mytok}{##1}%
{\color{eTagColor}\bfseries\listbreak}%
{}%
}%
{\color{commentColor}%
\expandafter\docsvlist\expandafter{\eTags}#1}%
}%

\lstdefinestyle{eclipse}{
  basicstyle={\lstfontfamily},
  emphstyle=\bfseries,
  keywordstyle=\color{keywordColor}\bfseries,
  commentstyle=\markupComments,
  stringstyle=\color{stringColor},
  numberstyle=\color{lineNumberColor}\lstfontfamily,
  morecomment=[s][\markupJavadocs]{/**}{*/}, 
  showstringspaces=false,
  numbers=left,
}

\lstdefinestyle{black}{
  basicstyle=\small\lstfontfamily,
  numbers=left,
  columns=fullflexible,
  breaklines=true,
  mathescape=true,
  escapechar=\#,
  tabsize=4,
  frame=lines,
  showstringspaces=false
}

\lstdefinestyle{seminar}{
  basicstyle=\small\ttfamily,
  numbers=left,
  breaklines=true,
  mathescape=true,
  escapechar=\#,
  tabsize=4,
  showstringspaces=false
}

\expandafter\lstset\expandafter{\jAnnotations}

\usepackage{tcolorbox}
\usepackage{pifont}
\usepackage{csquotes}
\usepackage{url}

\AtBeginDocument{%
  \providecommand\BibTeX{{%
    \normalfont B\kern-0.5em{\scshape i\kern-0.25em b}\kern-0.8em\TeX}}}

\setcopyright{acmcopyright}
\copyrightyear{2022}
\acmYear{2022}
\acmDOI{XXXXXX.XXXXXXX}

\begin{document}

\title{Selectively Combining Multiple Coverage Goals in Search-Based Unit Test Generation}


\author{Zhichao Zhou}
\affiliation{%
  \institution{School of Information Science and Technology,\\
  ShanghaiTech University}
  \state{Shanghai}
  \country{China}}
\email{zhouzhch@shanghaitech.edu.cn}

\author{Yuming Zhou}
\affiliation{%
  \institution{State Key Laboratory for Novel Software Technology,\\
  Nanjing University}
  \state{Nanjing}
  \country{China}}
\email{zhouyuming@nju.edu.cn}

\author{Chunrong Fang}
\affiliation{%
  \institution{State Key Laboratory for Novel Software Technology,\\
  Nanjing University}
  \state{Nanjing}
  \country{China}}
\email{fangchunrong@nju.edu.cn}

\author{Zhenyu Chen}
\affiliation{%
  \institution{State Key Laboratory for Novel Software Technology,\\
  Nanjing University}
  \state{Nanjing}
  \country{China}}
\email{zychen@nju.edu.cn}

\author{Yutian Tang}
\affiliation{%
  \institution{ShanghaiTech University}
  \state{Shanghai}
  \country{China}}
\email{csytang@ieee.org}
\authornote{Yutian Tang is the corresponding author.}



\begin{abstract}
Unit testing is a critical part of software development process, ensuring the correctness of basic programming units in a program (e.g., a method). Search-based software testing (SBST) is an automated approach to generating test cases. SBST generates test cases with genetic algorithms by specifying the coverage criterion (e.g., branch coverage). However, a good test suite must have different properties, which cannot be captured by using an individual coverage criterion. Therefore, the state-of-the-art approach combines multiple criteria to generate test cases. As combining multiple coverage criteria brings multiple objectives for optimization, it hurts the test suites' coverage for certain criteria compared with using the single criterion. To cope with this problem, we propose a novel approach named \textbf{smart selection}. Based on the coverage correlations among criteria and the coverage goals' subsumption relationships, smart selection selects a subset of coverage goals to reduce the number of optimization objectives and avoid missing any properties of all criteria. We conduct experiments to evaluate smart selection on $400$ Java classes with three state-of-the-art genetic algorithms. On average, smart selection outperforms combining all goals on $65.1\%$ of the classes having significant differences between the two approaches.
\end{abstract}

\ccsdesc[500]{Software and its engineering~Search-based software engineering}
\ccsdesc[500]{Software and its engineering~Software testing and debugging}

\keywords{SBST, software testing, test generation}

\maketitle

\section{Introduction} \label{sec:intro}


Unit testing is a common way to ensure software quality. Manually preparing unit tests can be a tedious and error-prone process. Hence, developers and researchers put much effort into automatically generating test cases for programming units over these years.

Search-based software testing (SBST) is considered a promising approach to generating test cases. It generates test cases with genetic algorithms (e.g. Whole Suite Generation (WS) \cite{FraserWhole}, MOSA \cite{PanichellaMOSA}, DynaMOSA \cite{PanichellaDynaMOSA}) by specifying the coverage criterion (e.g., branch coverage). The execution of a genetic algorithm relies on fitness functions, which quantify the degree to which a solution (i.e., one or more test cases) achieves its goals (i.e., satisfying a certain coverage criterion). For each coverage criterion, there is a group of fitness functions. Each fitness function describes whether or how far a test case covers a goal (e.g., a branch).

\begin{figure}[htbp]
    \centering
    \includegraphics[width=0.48\textwidth]{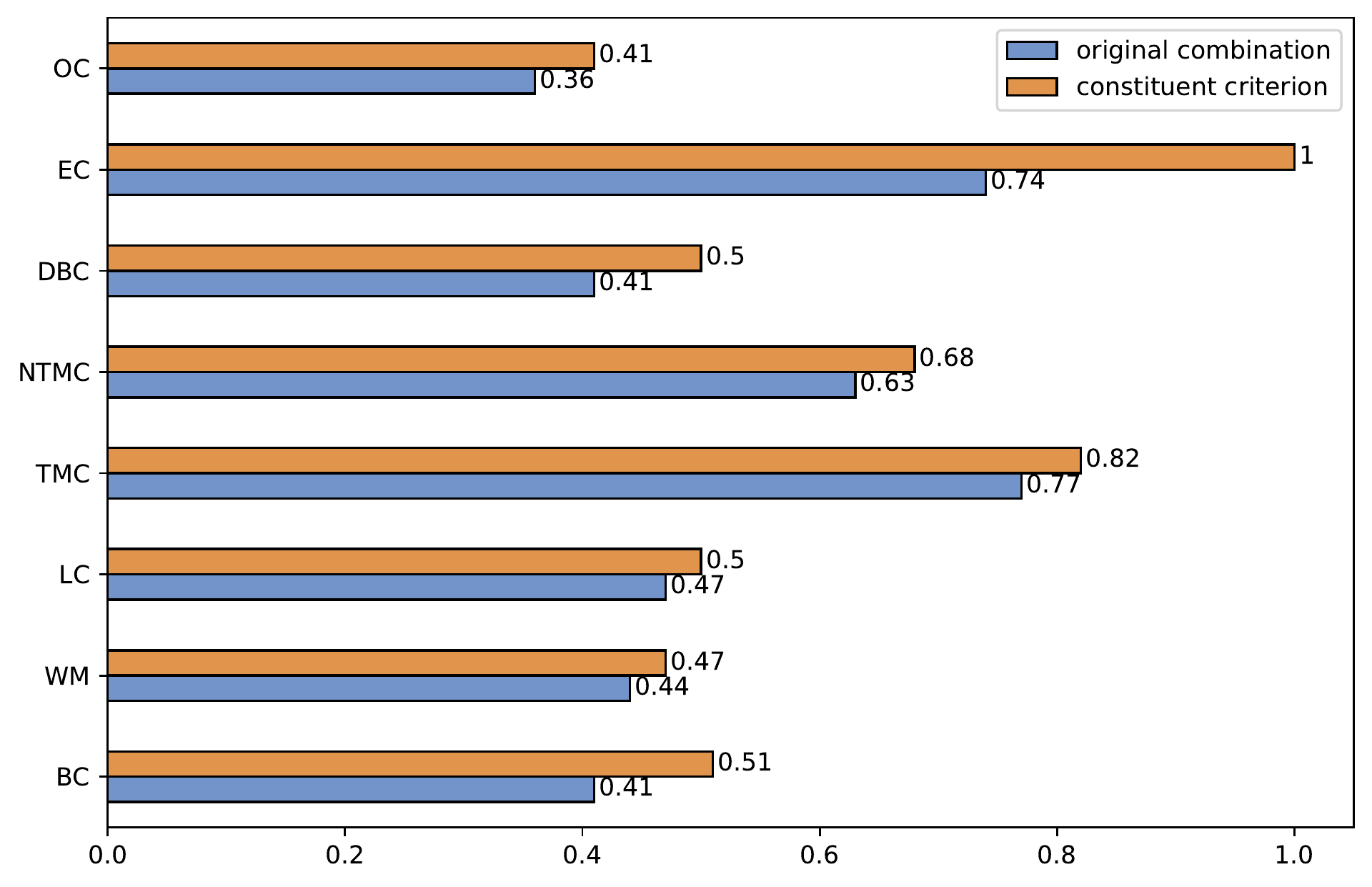}
    \caption{Partial Data of Coverage Gaps}
    \label{fig:coverage-gaps-1}
\end{figure}

\noindent\textbf{The Problem and Motivation.} However, as claimed in \cite{Rojas2015CombiningMC}, a good test suite must have different properties, which cannot easily be captured by any individual coverage criterion alone. Therefore, to generate a good test suite, multiple coverage criteria should be considered in SBST. Hence, the state-of-the-art approach \cite{Rojas2015CombiningMC} combines multiple coverage criteria to guide genetic algorithms. It involves eight coverage criteria (see Sec. \ref{subsec:criterion}). We call this method the \textbf{original combination} in this paper. However, combining multiple criteria brings more objectives for optimization, potentially affecting the effectiveness of the genetic algorithms \cite{KnowlesSingle, PanichellaMOSA, BrockhoffMulti}. For example, it can increase the probability of being trapped in local optima. As a result, the generated test suite's coverage decreases for certain criteria compared with using a single criterion. Fig. \ref{fig:coverage-gaps-1} shows (see Sec. \ref{subsec:rq1}) the average coverage gaps between the original combination and each constituent criterion when applying WS into $85$ big Java classes (i.e., with at least $200$ branches). The average gap of eight criteria is $8.2\%$. Branch coverage (BC) decreases $10\%$ and exception coverage (EC) decreases $26\%$. Note that since we cannot know the total exceptions in a class \cite{Rojas2015CombiningMC}, we normalize the exception coverage values of two approaches ($22.08$ vs. $29.74$) by dividing by the larger one.






%

\noindent\textbf{Targets.} To cope with this problem, a qualified approach should fulfil the following targets:
(1) \textbf{T1: GA Effectiveness.} It should select a subset from multiple coverage criteria' coverage goals. This subset improves the effectiveness of guiding genetic algorithms (GAs); and (2) \textbf{T2: Property Consistence.} This subset should avoid missing any properties captured by these coverage criteria.  

\noindent\textbf{Our Solution.}
To fulfil these targets, we propose a novel approach named \textbf{smart selection} (see Sec. \ref{sec:smart}). In this paper, we also consider the above eight coverage criteria. Instead of directly combining them, in smart selection, we firstly group them into four groups based on coverage correlations (see Sec. \ref{subsec:cc}). Next, we select one representative criterion that is more effective to guide the genetic algorithms from each group (T1) (see Sec. \ref{subsec:rc}). These selected coverage criteria ($SC$)' coverage goals are marked as $Goal(SC)$. To keep the property consistency  (T2), for each criterion ($c$) of unselected criteria ($USC$), we select a subset $Goal(c)_{sub}$ from its coverage goals based on the goals' subsumption relationships (see Sec. \ref{subsec:rs}). Finally, we combine $Goal(SC)$ and $\underset{c\in USC}{\cup}Goal(c)_{sub}$ to guide the test case generation process.



\noindent\textbf{Contribution.} In summary, the contribution of this paper includes:

\noindent$\bullet$ To the best of our knowledge, this is the \emph{first} paper that uses coverage correlations to address the coverage decrease caused by combining multiple criteria in SBST.

\noindent$\bullet$ We implement smart selection atop EvoSuite. It is integrated into three search algorithms (i.e., WS, MOSA, and DynaMOSA). 

\noindent$\bullet$ We conduct experiments on 400 Java classes to compare smart selection and the original combination. On average of three algorithms (WS/MOSA/DynaMOSA), smart selection outperforms the original combination on $77$ ($121$/$78$/$32$) classes, accounting for $65.1\%$ ($85.8\%$/$65\%$/$44.4\%$) of the classes having significant differences between the two approaches. The counterpart data of the $85$ big classes is $34$ ($50$/$35$/$16$), accounting for $86.1\%$ ($98\%$/$87.5\%$/$72.7\%$). Furthermore, we conduct experiments to compare smart selection with/without the subsumption strategy on $173$ classes.

\noindent\textbf{Online Artifact.} The online artifact of this paper can be found at: \url{https://doi.org/10.5281/zenodo.6467640}.
\section{Background}
\begin{figure}[htbp]
    \centering
    \includegraphics[width=0.5\textwidth]{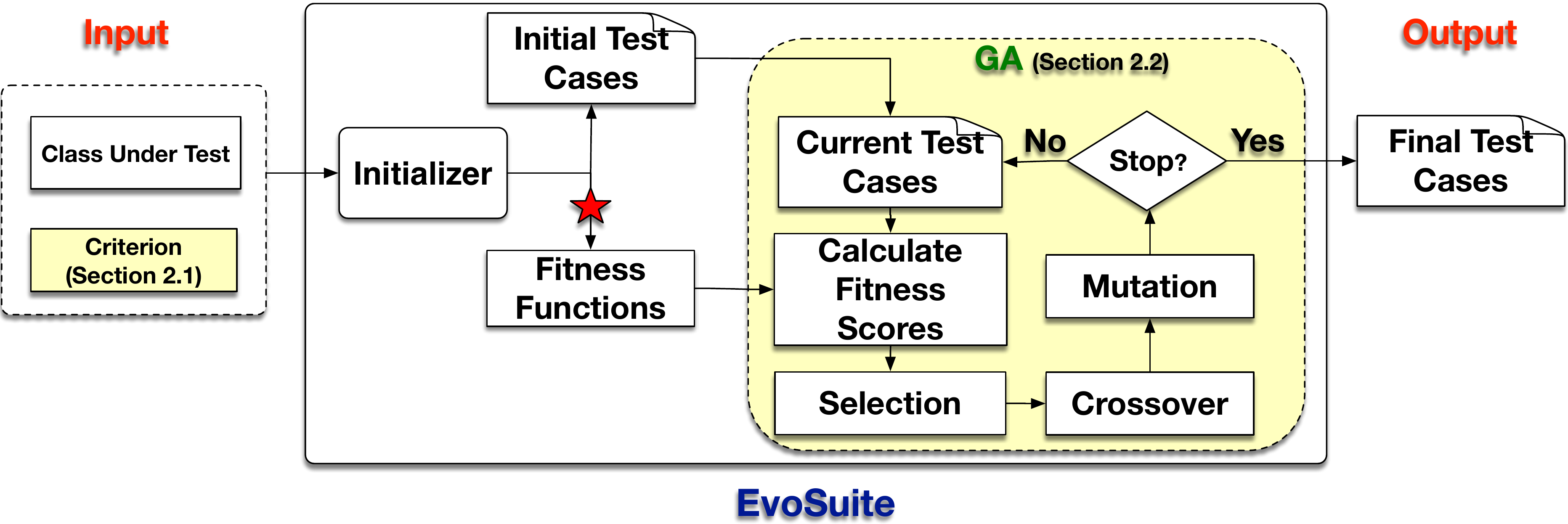}
    \caption{Overview of Unit Tests Generation in EvoSuite}
    \label{fig:evosuite}
\end{figure}
\begin{figure*}[t]
    \centering
    \includegraphics[width=\textwidth]{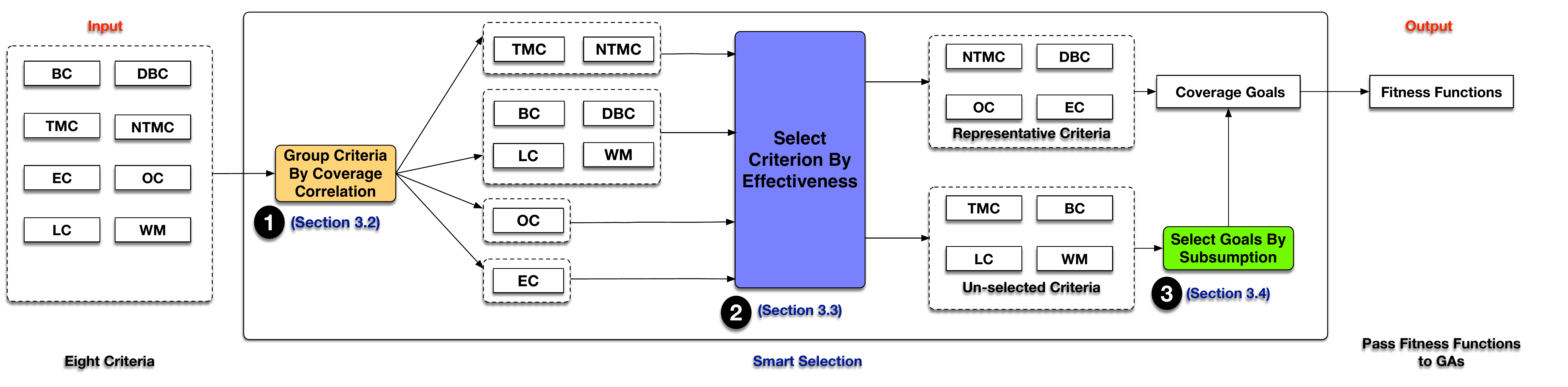}
    \caption{Overview of \textbf{Smart Selection}}
    \label{fig:ss}
\end{figure*}
\noindent\textbf{SBST and EvoSuite} \label{subsec:evosuite}
SBST generates test cases with a genetic algorithm (see Sec. \ref{subsec:ga}) by specifying the coverage criterion (see Sec. \ref{subsec:criterion}). EvoSuite \cite{FraserEvoSuite} is the state-of-art SBST tool for Java. In this section, we leverage EvoSuite as an example to illustrate the key idea of SBST. Fig. \ref{fig:evosuite} shows the overview of EvoSuite. The red star in this figure is mentioned later in Sec. \ref{subsec:ss-overview}.

\noindent\textbf{Input.}
Evosuite takes two major inputs: (1) the class under test (CUT) and (2) a coverage criterion (Sec. \ref{subsec:criterion}).

\noindent\textbf{Test Generation.} 
The test generation process can be divided into two parts: (1) The initializer extracts all the related information needed by the genetic algorithm (e.g., method signatures, including name, parameter type) from the CUT. Based on the information and the criterion, the initializer generates initial test cases and fitness functions. In general, each GA requires one or more specific fitness functions. A fitness function measures how close a test case covers a coverage goal (e.g., a branch); (2) After a specific genetic algorithm is invoked, it selects test cases based on the scores returned by fitness functions. Next, the GA creates new test cases with the crossover and mutation operations \cite{FraserWhole}. The GA repeats to select, mutate, and crossover test cases until all fitness functions reach the optima or the given budgets are consumed out.

\noindent\textbf{Output.}
After running the genetic algorithm, EvoSuite outputs the final test cases.

\subsection{Coverage Criteria}\label{subsec:criterion}
In this research, we discuss eight criteria as follows. The reason to choose these coverage criteria is that they are EvoSuite’s default criteria and are widely used in many previous studies \cite{Rojas2015CombiningMC, GayCombine, CamposCombineDefect}.

\noindent\textbf{Branch Coverage (BC)} 
BC checks the number of branches of conditional statements covered by a test suite. 

\noindent\textbf{Direct Branch Coverage (DBC)}
The only difference between DBC and BC is that a test case must directly invoke a public method to cover its branches. DBC treats others in the same way as BC \cite{Rojas2015CombiningMC}.

\noindent\textbf{Line Coverage (LC)}
LC checks the number of lines covered by a test suite.

\noindent\textbf{Weak Mutation (WM)}
WM checks how many mutants are detected by a test suite \cite{HarmanMut, petrovic2021mutation}. A mutant is a variant of the CUT generated by a mutation operator. For example, RC is an operator that replaces a constant value with different values \cite{FraserMutation}.

\noindent\textbf{Top-Level Method Coverage (TMC)}
TMC checks the number of methods covered by a test suite with a requirement: A public method is covered only when it is directly invoked by test cases.

\noindent\textbf{No-Exception Top-Level Method Coverage (NTMC)}
NTMC is TMC with an extra requirement: A method is not covered if it exits with any exceptions.

\noindent\textbf{Exception Coverage (EC)}
EC checks the number of exceptions triggered by a test suite.

\noindent\textbf{Output Coverage (OC)}
OC measures the diversity of the return values of a method. For example, a \textit{boolean} return variable's value can be \textit{true} or \textit{false}. OC's coverage is $100\%$ only if the test suite captures these two return values.

For each criterion, EvoSuite extracts a group of coverage goals from the CUT and assigns a fitness function to each coverage goal. For example, EvoSuite extracts all branches for BC, e.g., the true branch of a predicate $x==10$. A simplified fitness function of this branch can be the branch distance \cite{McMinnSbSurvey}, $|x-10|$, showing how far a test case covers it. More details of these criteria and their fitness functions can be found here \cite{Rojas2015CombiningMC}.

\subsection{Genetic Algorithms}\label{subsec:ga}
In this research, we discuss three GAs (i.e., WS, MOSA, DynaMOSA) as follows. All of them are integrated into EvoSuite and perform well in many SBST competitions \cite{vogl2021evosuite, Panichella2020evosuite, campos2019evosuite}. These algorithms share the same inputs and outputs but differ in how to use fitness functions.

\noindent\textbf{WS.}
WS \cite{FraserWhole} directly evolves test suites to fit all coverage goals. Consequently, WS can exploit the collateral coverage \cite{arcuri2011random} and not waste time on infeasible goals (e.g., dead code). The collateral coverage means that a test case generated for one goal can implicitly also cover any number of other coverage goals. Hence,  WS's fitness function is the sum of all goals' fitness functions.

\noindent\textbf{MOSA.}
WS sums fitness scores of all coverage goals as a scalar value. This scalar value is less monotonic and continuous than a single goal's fitness score, increasing the probability of being trapped in local optima. To overcome this limit, Panichella et al. \cite{PanichellaMOSA} formulates SBST as a many-objective optimization problem and propose MOSA, a variant of NSGA-\uppercase\expandafter{\romannumeral2} \cite{NSGA2}. In general, MOSA maintains a fitness vector for each test case. An item of the fitness vector is a fitness function value for the test case. Based on Pareto dominance  \cite{DebMulti}, MOSA sorts and selects test cases by the fitness vectors.

\noindent\textbf{DynaMOSA.}
Based on MOSA, DynaMOSA \cite{PanichellaDynaMOSA} adopts control dependency graph to reduce the coverage goals evolved in search. A goal is selected to be in the evolving process only when the branch goals it depends on are covered. Hence, the sizes of DynaMOSA's fitness vectors are often smaller than MOSA's ones. Empirical studies show that DynaMOSA outperforms WS and MOSA \cite{PanichellaDynaMOSA,CamposCombineDefect}.
\subsection{Combining Coverage Criteria}\label{subsec:combine}
With a single criterion (e.g., BC) alone, SBST can generate test cases that reach higher code coverage but fail to meet users' expectations \cite{Rojas2015CombiningMC}. Hence, Rojas et al. \cite{Rojas2015CombiningMC} proposed to combine multiple criteria to guide SBST to generate a test suite. We leverage replacing BC by combining the eight criteria as an example to show the changes in GAs. Before the combination, the fitness function of WS is $f_{BC}=\sum \limits_{b \in B}f_{b}$, where B is the set of all branches. The fitness vector of MOSA/DynaMOSA is $[f_{b_{1}},...,f_{b_{n}}].$ After the combination, the fitness function of WS is $f_{BC}+...+f_{OC}$. The fitness vector of MOSA/DynaMOSA is $[f_{b_{1}},...,f_{b_{n}},...,f_{o_{1}},...,f_{o_{m}}],$ where $o_{i}$ is a output coverage goal.
\section{Smart Selection}\label{sec:smart}
\textbf{The Problem and Motivation:} 
The main side-effect of combining multiple criteria is that the generated test suite's coverage decreases for certain criteria. It is due to the increase in optimization objectives. Firstly, It brings a \textit{larger} search space, reducing the search weight of each objective. Secondly, it also brings a \textit{harder} search space since some criteria' fitness functions are not monotonic and continuous, such as LC and WM (see Sec. \ref{subsec:rc}). Hence, we propose \textbf{smart selection}. It aims to relieve the coverage decrease by providing a smaller and easier search space for GAs.

\subsection{Overview} \label{subsec:ss-overview}
Fig. \ref{fig:ss} shows the process: \ding{182} group criteria by coverage correlation (Sec. \ref{subsec:cc}), \ding{183} select representative criteria by effectiveness to guide SBST ( Sec. \ref{subsec:rc}), and \ding{184} select representative coverage goals from unselected criteria by subsumption relationships (Sec. \ref{subsec:rs}). The red star in Fig. \ref{fig:evosuite} shows the position of smart selection in EvoSuite.

The inputs are the eight criteria (see Sec. \ref{subsec:criterion}). The output is a subset of fitness functions, i.e., the corresponding fitness functions of our selected coverage goals. This subset is used to guide GAs.

\subsection{Group Criteria by Coverage Correlation}\label{subsec:cc}
The first step is clustering these eight criteria. The standard of whether two criteria can be in one group or not is: whether these two criteria have a coverage correlation. Based on this standard, we can divide these criteria into several groups. For two criteria with coverage correlation, if a test suite achieves a high coverage under one of the criteria, then this test suite may also achieve a high coverage under the other criterion. Hence, we can select one of these two criteria to guide SBST. Thus, grouping sets the scope for choosing representative criteria.

We determine that two criteria have a coverage correlation by the following rules:

\noindent$\bullet$\textbf{Rule 1:} If a previous study shows that two criteria have a coverage correlation, we adopt the conclusion:

\noindent\ding{172} \textbf{BC and WM:} Gligoric et al. \cite{GligoricCorr} find that \enquote{branch coverage performs as well as or better than all other criteria studied, in terms of ability to predict mutation scores}. Their work shows that the average Kendall’s $\tau_{b}$ value \cite{kendall1938new} of coverage between branch coverage and mutation testing is $0.757$. Hence, we assume that BC and WM have a coverage correlation.

\noindent\ding{173} \textbf{DBC and WM:} Since BC and WM have a coverage correlation, we assume that DBC and WM have a coverage correlation too.

\noindent\ding{174} \textbf{LC and WM:} Gligoric et al. \cite{GligoricCorr} find that statement coverage \cite{GligoricCorr, Rojas2015CombiningMC} can be used to predict mutation scores too. Line coverage is an alternative for statement coverage in Java since Java's bytecode instructions may not directly map to source code statements \cite{Rojas2015CombiningMC}. Hence, we assume that LC and WM have a coverage correlation.

\noindent$\bullet$\textbf{Rule 2:} Two criteria, A and B, have a coverage correlation if they satisfy two conditions: (1) A and B have the same coverage goals; (2) A test covers a goal of A only if it covers the counterpart goal of B and it satisfies A's additional requirements:

\noindent\ding{175} \textbf{DBC and BC:} DBC is BC with an additional requirement (see Sec. \ref{subsec:criterion}).

\noindent\ding{176} \textbf{NTMC and TMC:} NTMC is TMC with an additional requirement (see Sec. \ref{subsec:criterion}).

\noindent$\bullet$\textbf{Rule 3:} Two criteria, A and B, have a coverage correlation if, for an arbitrary test suite, the relationship between these two criteria' coverage (i.e., $C_{A}$ and $C_{B}$) can be formulated as:
    \begin{equation}
    C_{B} = \Theta C_{A},
    \end{equation}
where $\Theta$ is a nonnegative random variable and $E\Theta \approx 1$:

\noindent\ding{177} \textbf{BC and LC:} Intuitively, when a branch is covered, then all lines in that branch are covered. But this is not always true. When a line exits abnormally (e.g., it throws an exception.), the subsequent lines are not covered either. First, we discuss the coverage correlation of branch and line coverage in the absence of abnormal exiting. Let $B$ be the set of branches of the CUT, $L$ be the set of lines, and $T$ be a test suite. For any $b \in B$, let $L_{b}$ be the set of lines \textbf{only} in the branch $b$ (i.e., we don't count the lines in its nested branches). Consequently, $L=\bigcup_{b \in B} L_{b}$. Let $B^{'}$ be the set of covered branches. Let $L^{'}$ be the set of covered lines. The coverage values measured by branch and line coverage are:
    \begin{align}
        C_{Branch} = \frac{|B^{'}|}{|B|}, 
        C_{Line} = \frac{|L^{'}|}{|L|} = \frac{\sum \limits_{b \in B^{'}}|L_{b}|}{\sum \limits_{b \in B}|L_{b}|}.
    \end{align}
    Hence, the relationship of $C_{Branch}$ and $C_{Line}$ is:
    \begin{align}
        \frac{C_{Line}}{C_{Branch}} & = \frac{\sum \limits_{b \in B^{'}}|L_{b}|}{\sum \limits_{b \in B}|L_{b}|} \div \frac{|B^{'}|}{|B|} =  \frac{\sum \limits_{b \in B^{'}}|L_{b}|}{|B^{'}|} \div \frac{\sum \limits_{b \in B}|L_{b}|}{|B|}.
    \end{align}
    Suppose we treat branches with different numbers of lines equally in generating $T$. Then we have:
    \begin{equation}
     \frac{\sum \limits_{b \in B^{'}}|L_{b}|}{|B^{'}|} \approx \frac{\sum \limits_{b \in B}|L_{b}|}{|B|}, 
    \end{equation}
    i.e., 
    \begin{align}
          \frac{C_{Line}}{C_{Branch}} \approx 1.
    \end{align}\label{align:bl}
    As a result, branch coverage and line coverage have a coverage correlation in the absence of abnormal exiting. With abnormal exiting, the coverage measured by line coverage decreases. Assuming that any line can exit abnormally, we can formulate the coverage relationship as:
    \begin{align}
          C_{Line} = \Theta C_{Branch},
    \end{align}
    where $\Theta$ is a random variable. In this research, instead of analyzing $\Theta$ precisely, we only need to check whether $E\Theta \approx 1$. Previous work \cite{Rojas2015CombiningMC} shows that, on average, when $78\%$ of branches are covered,  test suites can only find $1.75$ exceptions. Hence, we assume that $E\Theta \approx 1$, i.e., BC and LC have a coverage correlation.

\noindent\ding{178} \textbf{DBC and LC:} We assume that DBC and LC have a coverage correlation since BC and LC have a coverage correlation.

\noindent\textbf{Output.} We cluster the eight criteria into four groups: (1) BC, DBC, LC, and WM; (2) TMC and NTMC; (3) EC; and  (4) OC.
\subsection{Select Representative Criterion by Effectiveness to guide SBST}\label{subsec:rc}
In this step,  among the criteria in each group, we select a criterion to represent the others. The criteria within a group differ in the ability to guide SBST. If we only select one criterion with the best effectiveness to guide SBST, SBST will be more efficient in generating unit tests. To select the best criterion to guide SBST in each group, we need to compare the criteria' effectiveness in guiding SBST. A criterion's effectiveness in guiding SBST largely depends on the continuity of monotonicity of its fitness functions \cite{LinGraph, lin2020recovering}. Hence, we need to analyze and compare the criteria' fitness functions.

\noindent\textbf{Group1: BC, DBC, LC and WM.} 
We use branch coverage as the baseline and divide them into three pairs for discussion. The reason to use branch coverage as the baseline is that branch coverage has been widely used to guide unit test generation \cite{FraserWhole, PanichellaMOSA, PanichellaDynaMOSA} due to the monotonic continuity of its fitness functions. For a branch goal $b$ and a test case $t$,  its fitness function is \cite{FraserWhole}:
\begin{equation}
    f_{bc}(b, t)=\left\{
\begin{array}{lcl}
    0 & & {\textnormal{if~the~branch}} \\
    & & {\textnormal{has~been~covered,}} \\
    \nu(d(b, t)) & & {\textnormal{if~the~predicate~has~been}} \\
    & & {\textnormal{executed~at~least~twice,}} \\
    1 & & {\textnormal{otherwise,}}
\end{array}\right.\label{fitness:branch}
\end{equation}
where $\nu(x)$ is a normalizing function in $[0,1]$ (e.g., $\nu(x)=x/(x+1)$). $d(b,t)$ is a function to provide a branch distance to describe how far a test case covers this goal \cite{McMinnSbSurvey}. To avoid an oscillate situation of a predicate \cite{FraserWhole}, $f_{bc}(b,t)$ uses $\nu(d(b, t))$ only when a predicate is executed at least twice.

WS uses the sum of all fitness functions as one fitness function (Sec. \ref{subsec:ga}). Hence, for WS, branch coverage's fitness function is:
\begin{equation}
    d_{1}(b,T) = \min{\{f_{bc}(b,t)|t\in T\}}, 
\end{equation}
\begin{equation}
    f_{BC}(T) = \sum_{b\in B}{d_{1}(b, T)}\label{fitness:BC},
\end{equation}
where $B$ denotes all branches of the CUT.

\noindent$\bullet$\textbf{BC vs. LC.}
Based on line coverage's definition (see Sec. \ref{subsec:criterion}), a line $l$'s fitness function can be:
\begin{equation}
f_{lc}(l, t) = \left\{\
\begin{array}{ll}
    0 &  {\textnormal{if~the~line~has~been}} \\
    &  {\textnormal{covered,}} \\
     1  &{\textnormal{otherwise.}} 
\end{array}
\right.\label{fitness:origin-line}
\end{equation}
For WS, line coverage's fitness function is:
\begin{equation}
    f_{LC}(T) = \nu(|L|-|CL|)\label{fitness:origin-LC},
\end{equation}
where $L$ is the set of all lines and $CL$ is the set of covered lines. 

These two fitness functions are not continuous and monotonic since they only tell whether the lines are covered. To overcome this limit, EvoSuite uses branch coverage's fitness functions to augment line coverage's fitness functions \cite{Rojas2015CombiningMC}. A line $l$'s fitness function is:
\begin{equation}
f_{lc}(l, B, t) = \left\{\
\begin{array}{ll}
    0 &  {\textnormal{if~the~line~has~been}} \\

    &  {\textnormal{covered,}} \\
     1 +\min{\{f_{bc}(b, t)|b \in B\}}  &{\textnormal{otherwise,}} 
\end{array}
\right.\label{fitness:line}
\end{equation}
where $B$ is the set of branches that $l$ depends on \cite{PanichellaDynaMOSA}. For WS, line coverage's fitness function is:
\begin{equation}
    f_{LC}(T) = \nu(|L|-|CL|) + f_{BC}(T).\label{fitness:LC}
\end{equation}

We call Equation \ref{fitness:origin-line} and \ref{fitness:origin-LC} \textit{def-based} (definition-based) fitness functions and call Equation \ref{fitness:line} and \ref{fitness:LC} \textit{augmented} fitness functions.

Firstly, we compare branch coverage's fitness functions with line coverage's \textit{def-based} fitness functions.  Line coverage's \textit{def-based} fitness functions are not continuous and monotonic since they only tell whether the lines are covered.  Hence, branch coverage is better than line coverage in the effectiveness to guide SBST when we use line coverage's \textit{def-based} fitness functions. After the augmentation, line coverage's \textit{def-based} fitness functions disturb the continuity and monotonicity of branch coverage's fitness functions, undermining branch coverage's effectiveness to guide SBST. As a result, BC is better than LC in the effectiveness to guide SBST.

\noindent$\bullet$\textbf{BC vs. WM.}
Based on weak mutation's definition (see Sec. \ref{subsec:criterion}), a mutant's fitness function is:
\begin{equation}
f_{wm}(\mu, t) = \left\{\
\begin{array}{ll}
   1 & {\textnormal{if~mutant~$\mu$}} \\
    & {\textnormal{was not reached,}} \\
    \nu(id(\mu, t)) & {\textnormal{if~mutant~$\mu$}} \\
    & {\textnormal{was~reached,}}
\end{array}\right.\label{fitness:origin-wm}
\end{equation}
where $id(\mu, t)$ is the infection distance function. It describes how distantly a test case triggers a mutant's different state from the source code. Different mutation operators have different infection distance functions \cite{FraserMutation}. A mutant's fitness function is always $1$ unless a test case reaches it (i.e., the mutated line is covered). Hence, like line coverage, EvoSuite uses the same way to augment weak mutation's fitness functions \cite{PanichellaDynaMOSA, FraserMutation}. As the conclusion of comparing BC and LC, BC is better than WM in the effectiveness to guide SBST.

\noindent$\bullet$\textbf{BC vs. DBC.}
Direct branch coverage is branch coverage with an extra requirement: A test case must directly invoke a public method to cover its branches. Based on branch coverage's fitness function, we can get direct branch coverage's one: For a branch in a public method, when the method is not invoked directly, the fitness function always returns 1. Otherwise, the fitness function is the same as branch coverage's one. It is easy for SBST to generate a test case that invokes a public method directly. Hence, we regard that BC is nearly equal to DBC in guiding SBST.

\noindent\textbf{Order of Group1.}
Above all, we get a rough order of this group: (1)BC and DBC; (2) LC and WM. Since we only need one representative, the rough order satisfies our need.

\noindent\textbf{The Representative Criterion of Group1.}
We choose DBC to represent this group instead of BC. The reason is:  When a test case covers a goal of DBC, the test case covers the counterpart of BC. As a result, DBC can fully represent BC. The opposite may not hold.

\noindent\textbf{Group2: TMC and NTMC.} 
Like the relationship between branch coverage and direct branch coverage, no-exce. top-level method coverage is top-level method coverage with an extra requirement: A method must be invoked without triggering exceptions.

\noindent\textbf{The Representative Criterion of Group2.}
We choose NTMC to represent this group. The reason is the same as why we choose DBC to represent group 1: NTMC can fully represent TMC. The opposite does not hold.

\noindent\textbf{Group3: EC and Group 4: OC.} 
Since group 3 only contains EC, we choose EC to represent group 3. Similarly, we choose OC to represent group 4.

\noindent\textbf{Output.}
The representative criteria are DBC, NTMC, EC, and OC. 
\subsection{Select Representative Coverage Goals by Subsumption Relationships}\label{subsec:rs}
After selecting the representative criteria in the previous step, there are four unselected criteria: LC, WM, BC, and TMC. To keep property consistency for each unselected criterion, we select a subset from its coverage goals. This subset can represent all properties required by this criterion, ensuring GA archives \cite{PanichellaDynaMOSA} those tests that fulfill the properties beyond the representative criteria. We have another requirement for these subsets: they should be as small as possible. These unselected criteria' fitness functions are less continuous and monotonic than the ones of the representative criteria (see Sec. \ref{subsec:rc}). Hence, to minimize the negative effects on guiding SBST, these subsets should be as small as possible.

Two coverage goals, $G_{1}$ and $G_{2}$, having the subsumption relationship denotes that if a test suite covers one coverage goal, it must cover another goal. Specifically, $G_{1}$ subsuming $G_{2}$ represents that if a test suite covers $G_{1}$, it must cover $G_{2}$. According to this definition, for a criterion, if the coverage goals not subsumed by others are covered, all coverage goals are covered. Hence, These coverage goals form the desired subset.

\noindent\textbf{LC.}
For the lines in a basic block, the last line subsumes others. Hence, these last lines of all basic blocks form the desired subset. Since Sec. \ref{subsec:cc} shows that BC/DBC and LC have a strong coverage correlation and DBC is the representative criterion, we do a tradeoff to shrink this subset: We add an integer parameter \textit{lineThreshold}. If a basic block's lines are less than \textit{lineThreshold}, we skip it. In this research, we set \textit{lineThreshold} as $8$ (Sec. \ref{subsec:pt} discusses it).

\noindent\textbf{WM.}
The process to extract the subset from weak mutation's all mutants can be divided into three parts: \noindent\ding{172} We select the key operators from EvoSuite's all implemented mutation operators; \noindent\ding{173} From the key operators we filter out the \textbf{equal-to-line} operators; \noindent\ding{174} For the remaining operators, we select the subsuming mutants by following the previous work \cite{GheyiSub}.
  
\noindent\ding{172} \textbf{Select Key Operators.}
Offutt et al. \cite{OffuttOperators} find that five key operators achieve $99.5\%$ mutation score. They are UOI, AOR, ROR, ABS, and LCR. EvoSuite does not implement LCR (an operator that replaces the logical connectors) and ABS (an operator that inserts absolute values) \cite{FraserMutation}. Hence, we select three operators: UOI, AOR, and ROR (see Table \ref{tab:mo}).
 \begin{table}[!htpb]
 \centering
 \caption{\textbf{Evosuite}'s Partial Mutation Operators}\label{tab:mo}
 \begin{tabular}
 {@{\extracolsep{\fill}}|l|l|}
 \hline
    Operator  & Usage  \\ \hline
     UOI  & Insert unary operator \\ \hline
     AOR & Replace an arithmetic operator \\ \hline
     ROR & Replace a comparison operator \\ \hline
 \end{tabular}
  \end{table}
  \begin{figure*}[t]
\centering
\includegraphics[width=1\textwidth]{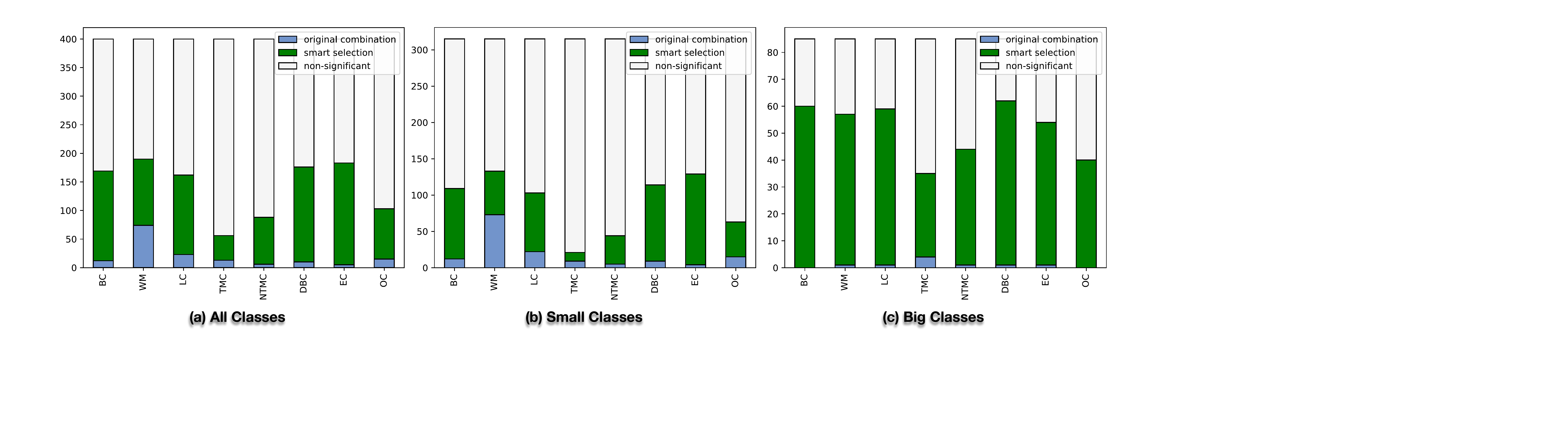}
\caption{Significant case summary of smart selection and the original combination with WS}\label{fig:suite_os}
\end{figure*}
  
\noindent\ding{173} \textbf{Filter out Equal-To-Line Operators.}
For each mutation operator, EvoSuite designs an infection distance function to describe how far a mutant's different state from the source code is triggered \cite{FraserMutation}. Some infection distance functions always return $0$. For example, UOI only adds $1$ to, subtracts $1$ to, or negates a numerical value, so the infection distance is always $0$. Hence, if a test case covers the line mutated by UOI, the mutant is killed. We call this kind of operator an \textbf{equal-to-line} operator. Among three key operators, only UOI is the equal-to-line operator \cite{FraserMutation}. Since line coverage has been dealt with, we filter out it.

\noindent\ding{174} \textbf{Select Subsuming Mutants.}
The remains are AOR and ROR. We choose one of the existing approaches \cite{mresa1999efficiency,JiaHOM,JustRedundant,GdynaMutant,GheyiSub} to select subsuming mutants for them. These approaches can be divided into three categories: (1) Manual analysis: Just et al. \cite{JustRedundant} build the subsumption relationships for ROR and LCR by analyzing all possible outputs of their mutants. This approach can not be applied to non-logical operators \cite{GdynaMutant}; (2) Dynamic analysis: By running an exhaustive set of tests, Guimar{\~a}es et al. \cite{GdynaMutant} build the subsumption relationships. This approach needs many tests, which we can not provide; (3) Static analysis: Gheyi and Souza et al. \cite{GheyiSub, souza2020identifying} encode a theory of subsumption relations in the Z3 theorem prover to identify the subsumption relationships. We adopt this approach because (\textrm{i}) This approach can be applied to both AOR and ROR; (\textrm{ii}) Using the Z3 prover to identify the subsumption relationships is a once-for-all job. We can hardcode their results into EvoSuite.

\noindent\textbf{BC and TMC.}
For a coverage goal of branch coverage, there is a subsuming goal from direct branch coverage (see Sec. \ref{subsec:cc}). As a result, the subset for branch coverage is empty since we select direct branch coverage as the representative (see Sec. \ref{subsec:rc}). Similarly, the subset for top-level method coverage is empty too.

\noindent\textbf{Output.}
For four unselected criteria, we select four subsets of their coverage goals. Two of them are empty. Finally, smart selection joins these subsets with the representative criteria to get their fitness functions for guiding GAs.
\section{Evaluation}
\subsection{Experiment Setting}\label{subsec:exp_set}
The evaluation focuses on the performance of smart selection. Our evaluation aims to answer the following research questions:

\noindent$\bullet$\textbf{RQ1:} How does smart selection perform with WS?

\noindent$\bullet$\textbf{RQ2:} How does smart selection perform with MOSA?

\noindent$\bullet$\textbf{RQ3:} How does smart selection perform with DynaMOSA?

\noindent$\bullet$\textbf{RQ4:} How does the subsumption strategy affect the performance of smart selection?

\noindent\textbf{Environment.} All experiments are conducted on two machines with Intel(R) Core(TM) i9-10900 CPU @ 2.80GHz and 128 GB RAM.

\noindent\textbf{Subjects.} 
We randomly select Java classes from $2$ sources: the benchmark of DynaMOSA \cite{PanichellaDynaMOSA} and Hadoop \cite{hadoop}. Following the previous work \cite{PanichellaMOSA}, the only restriction of randomly selecting classes is that the class must contain at least $50$ branches, aiming to filter out the trivial classes. As a result, we select $400$ classes: $158$ from the benchmark of DynaMOSA and $242$ from Hadoop.
 
\noindent\textbf{Baseline for RQ1-3.} We have two baselines: (1) the original combination, used to be compared with smart selection on each Java class; (2) a single constituent criterion, used to show the data of coverage decrease caused by the above two combination approaches.  A single constituent criterion means that we only use each criterion of these eight criteria (see Sec. \ref{subsec:criterion}) to guide GAs. There is one exception: when the constituent criterion is exception or output coverage, we combine this criterion and branch coverage to guide GAs. The reason is that only exception or output coverage is weak in the effectiveness of guiding the GAs \cite{Rojas2015CombiningMC, GayCombine}. Branch coverage can guide the GAs to reach more source lines of the CUT, increasing the possibility of triggering exceptions or covering output goals.

\noindent\textbf{Configuration for RQ1-3.}
EvoSuite provides many parameters (e.g., crossover probability, population size \cite{FraserWhole}) to run the algorithms. In this paper, we adopt EvoSuite's default parameters to run smart selection and other baselines. 

Smart selection introduces a new parameter \textit{lineThreshold} (see Sec. \ref{subsec:rs}). It controls smart selection to skip basic blocks with less than \textit{lineThreshold} lines. We set  \textit{lineThreshold} as $8$. The discussion on this value is in Sec. \ref{subsec:pt}. For each Java class, we run EvoSuite with ten approaches: (1) smart selection, (2) the original combination, and (3) each constituent criterion of all eight criteria. We run each approach for $30$ rounds per Java class, and each run's search budget is $2$ minutes.

\subsection{RQ1: How does smart selection perform with WS?}\label{subsec:rq1}
\noindent\textbf{Motivation.}
In this RQ, we evaluate smart selection (\textbf{SS}) with WS. First, we compare the performance of SS and the original combination (\textbf{OC}). Next, we use the coverage of each constituent criterion (\textbf{CC}) to show the coverage decrease caused by SS and OC. Furthermore, we show these approaches' differences in the resulted suite sizes (i.e., the number of tests in a test suite).

\noindent\textbf{Methodology.} 
EvoSuite records the coverage for generated unit tests. For each class, we obtain $10$ coverage data sets: One data set records the coverage of the eight criteria when using SS; One data set records the coverage of the eight criteria when using OC; The rest data sets record the coverage when using each CC.

For each Java class, we follow previous research work \cite{Rojas2015CombiningMC} to use Mann-Whitney U Test to measure the statistical difference between SS and OC. Then, we use the Vargha-Delaney $\hat{A}_{ab}$ \cite{vargha2000critique} to evaluate whether a particular approach $a$ outperforms another approach $b$ ($A_{ab}>0.5$ and the significant value $p$ is smaller than $0.05$).

\noindent\textbf{Result.}
\begin{table}[htbp]
	\centering
	\footnotesize
	\caption{Average coverage results for each approach with WS}
	\label{tab:suite}
\footnotesize{\textbf{(a)} All Classes}\\
\begin{tabular}{l| l| l |l  }
\hline
approach & SS & OC & CC \\ \hline
BC & 55\% & 53\% & \textcolor[RGB]{0,128,28}{57\%} \\ \hline
WM & \textcolor[RGB]{0,128,28}{59\%} & 57\% & \textcolor[RGB]{0,128,28}{59\%} \\ \hline
LC & \textcolor[RGB]{0,128,28}{60\%} & 58\% & \textcolor[RGB]{0,128,28}{60\%} \\ \hline
TMC & \textcolor[RGB]{0,128,28}{84\%} & 83\% & \textcolor[RGB]{0,128,28}{84\%} \\ \hline
\end{tabular}
\begin{tabular}{l| l| l |l  }
\hline
approach & SS & OC & CC \\ \hline
NTMC & \textcolor[RGB]{0,128,28}{71\%} & 70\% & \textcolor[RGB]{0,128,28}{71\%} \\ \hline
DBC & 55\% & 53\% & \textcolor[RGB]{0,128,28}{56\%} \\ \hline
EC & 15.92 & 14.52 & \textcolor[RGB]{0,128,28}{16.52} \\ \hline
OC & 44\% & 43\% & \textcolor[RGB]{0,128,28}{45\%} \\ \hline
\end{tabular}\\
\footnotesize{\textbf{(b)} Small Classes}\\
\begin{tabular}{l| l| l |l  }
\hline
approach & SS & OC & CC \\ \hline
BC & 58\% & 57\% & \textcolor[RGB]{0,128,28}{59\%} \\ \hline
WM & \textcolor[RGB]{0,128,28}{62\%} & 61\% & \textcolor[RGB]{0,128,28}{62\%} \\ \hline
LC & \textcolor[RGB]{0,128,28}{62\%} & 61\% & \textcolor[RGB]{0,128,28}{62\%} \\ \hline
TMC & 85\% & 85\% & 85\% \\ \hline
\end{tabular}
\begin{tabular}{l| l| l |l  }
\hline
approach & SS & OC & CC \\ \hline
NTMC & \textcolor[RGB]{0,128,28}{73\%} & 72\% & 72\% \\ \hline
DBC & 57\% & 56\% & \textcolor[RGB]{0,128,28}{58\%} \\ \hline
EC & \textcolor[RGB]{0,128,28}{13.29} & 12.48 & 12.95 \\ \hline
OC & \textcolor[RGB]{0,128,28}{46\%} & 45\% & \textcolor[RGB]{0,128,28}{46\%} \\ \hline
\end{tabular}\\
\footnotesize{\textbf{(c)} Big Classes}\\
\begin{tabular}{l| l| l |l  }
\hline
approach & SS & OC & CC \\ \hline
BC & 45\% & 41\% & \textcolor[RGB]{0,128,28}{51\%} \\ \hline
WM & \textcolor[RGB]{0,128,28}{48\%} & 44\% & 47\% \\ \hline
LC & \textcolor[RGB]{0,128,28}{50\%} & 47\% & \textcolor[RGB]{0,128,28}{50\%} \\ \hline
TMC & 79\% & 77\% & \textcolor[RGB]{0,128,28}{82\%} \\ \hline
\end{tabular}
\begin{tabular}{l| l| l |l  }
\hline
approach & SS & OC & CC \\ \hline
NTMC & 67\% & 63\% & \textcolor[RGB]{0,128,28}{68\%} \\ \hline
DBC & 45\% & 41\% & \textcolor[RGB]{0,128,28}{50\%} \\ \hline
EC & 25.69 & 22.08 & \textcolor[RGB]{0,128,28}{29.74} \\ \hline
OC & 39\% & 36\% & \textcolor[RGB]{0,128,28}{41\%} \\ \hline
\end{tabular}
\end{table}
\begin{table}[htbp]
    \centering
    \small
    \caption{Average test suite size of each approach with WS}
    \begin{tabular}{l|l|l|l}
    \hline
        approach & SS & OC & CC (Average)  \\ \hline
        size (All Classes) & 51.35 & 47.77 & 31.59 \\ \hline
        size (Small Classes) & 37.27 & 36.39 & 19.43 \\ \hline
        size (Big Classes) & 103.53 & 89.95 & 76.64 \\ \hline
    \end{tabular}
    \label{suite_size_ws}
\end{table}
\begin{figure*}[t]
\centering
\includegraphics[width=1\textwidth]{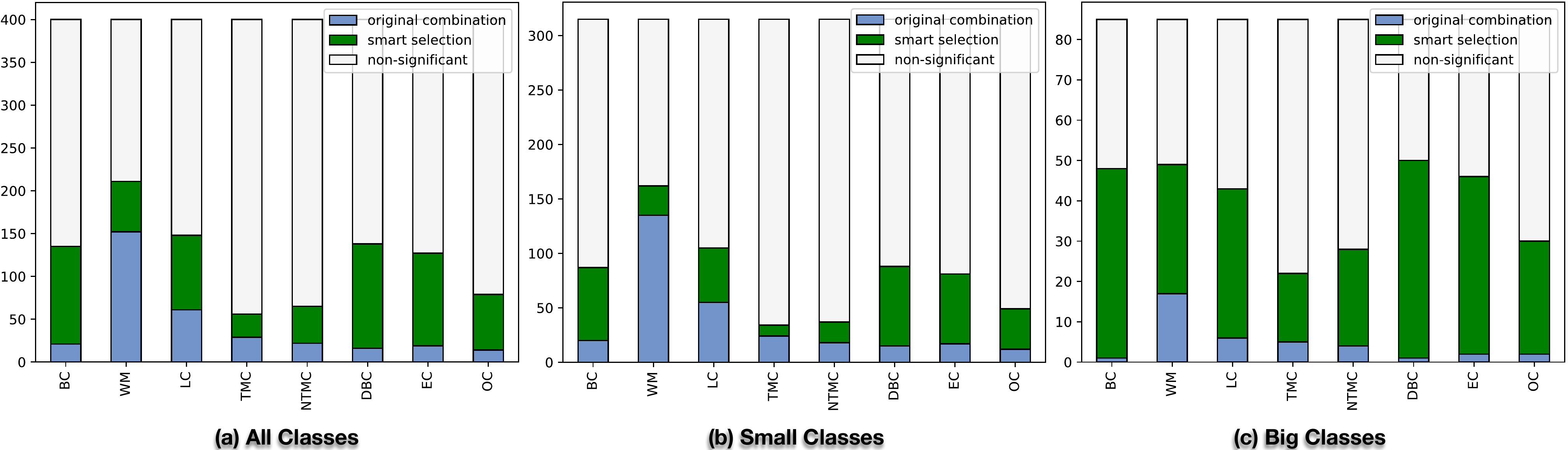}
\caption{Significant case summary of smart selection and the original combination with MOSA}\label{fig:mosa_os}
\end{figure*}
\noindent$\blacktriangleright$\textbf{All Classes.}

\noindent\textbf{Significant Cases.} Fig. \ref{fig:suite_os} (a) shows the comparison results of SS and OC on all $400$ Java classes. SS outperforms OC on $121$ ($30.3\%$) classes (a.k.a., SS-outperforming classes) on average for each coverage. OC outperforms SS on $20$ ($4.9\%$) classes (a.k.a., OC-outperforming classes). These two approaches have no significant difference on $259$ ($64.8\%$) classes (a.k.a., No-significant classes) on average.

\noindent\textbf{Average Coverage.} Table \ref{tab:suite} (a) shows the average coverage of all classes with three approaches. For exception coverage, the table shows the average number of the triggered exceptions since we can not know the total number of exceptions in a class \cite{Rojas2015CombiningMC}. The green number represents the highest coverage at a given criterion. SS outperforms OC for eight criteria' coverage. Among three approaches, SS reaches the highest coverage for four criteria. CC reaches the highest coverage for all criteria.

\noindent\textbf{Average Suite Size.} The first row of Table \ref{suite_size_ws} shows the test suites' average sizes of all classes. Compared to CC (average suite size of all constituent criteria), the size of OC increases by $51.2\%$ ($(47.77-31.59)/31.59$). Compared to OC, the size of SS increases by $7.4\%$ ($(51.35-47.77)/47.77$).

\noindent$\blacktriangleright$\textbf{Small Classes.($<$ 200 branches)}

\noindent\textbf{Significant Cases.} Fig. \ref{fig:suite_os} (b) shows the comparison results of SS and OC on $315$ small Java classes. For each criterion, on average, SS-outperforming classes are $71$ ($22.5\%$). OC-outperforming classes are $19$ ($5.9\%$). No-significant classes are $226$ ($71.6\%$).

\noindent\textbf{Average Coverage.} Table \ref{tab:suite} (b) shows the average coverage of small classes. SS outperforms OC for seven criteria' coverage. SS reaches the highest coverage for five criteria.

\noindent\textbf{Average Suite Size.} The second row of Table \ref{suite_size_ws} shows the average suite sizes of small classes. Compared to CC, the size of OC increases by $87.3\%$. Compared to OC, the size of SS increases by $2.4\%$.

\noindent$\blacktriangleright$\textbf{Big Classes. ($\geq$ 200 branches)}

\noindent\textbf{Significant Cases.} Fig. \ref{fig:suite_os} (c) shows the comparison results of SS and OC on $85$ big Java classes. For each criterion, on average, SS-outperforming classes are $50$ ($59.1\%$). The number of OC-outperforming classes is $1$ ($1.3\%$). No-significant classes are $34$ ($39.6\%$).
 
\noindent\textbf{Average Coverage.} Table \ref{tab:suite} (c) shows the average coverage of big classes. SS outperforms OC for eight criteria' coverage. SS reaches the highest coverage for two criteria.

\noindent\textbf{Average Suite Size.} The third row of Table \ref{suite_size_ws} shows the average suite sizes of big classes. Compared to CC, the size of OC increases by $17.4\%$. Compared to OC, the size of SS increases by $15.1\%$.

\noindent\textbf{Analysis.}
SS outperforms OC statistically, especially on the big classes. There is one exception: On the small classes, the number ($73$) of OC-outperforming classes in weak mutation is more than the counterpart number ($60$) (see Fig. \ref{fig:suite_os} (b)). On average, each of those $73$ classes has $82$ branches and $321$ mutants, while each of those $60$ classes has $115$ branches and $381$ mutants. It also supports that SS outperforms OC on the big classes. Furthermore, in most cases, the average coverage of CC is higher than the one of OC. SS narrows the coverage gap between them. For example, the biggest gap is $10\%$, happening in the big classes' branch coverage. SS narrows the gap by $4\%$ (see Table \ref{tab:suite} (c)). These facts confirm that combining criteria offers more objectives for optimization, affecting the efficacy of GAs. The bigger classes bring more objectives, leading to a higher impact. The suite size increase brought by OC/SS to CC is significant, confirming the experimental results of the work proposing OC \cite{Rojas2015CombiningMC}. The main reason is that the GA (not only WS but also MOSA/DynaMOSA) needs more tests for more goals. With the coverage increase, the suite size of SS is also greater than OC. Compared with the extent of the suite increase brought by OC to CC, we regard that it is reasonable.
\begin{tcolorbox}[title=Answer to RQ1,boxrule=1pt,boxsep=1pt,left=2pt,right=2pt,top=2pt,bottom=2pt]
With WS, SS outperforms OC statistically, especially on the big classes (i.e., the classes with no less than 200 branches).
\end{tcolorbox} 
\begin{figure*}[t]
\centering
\includegraphics[width=1\textwidth]{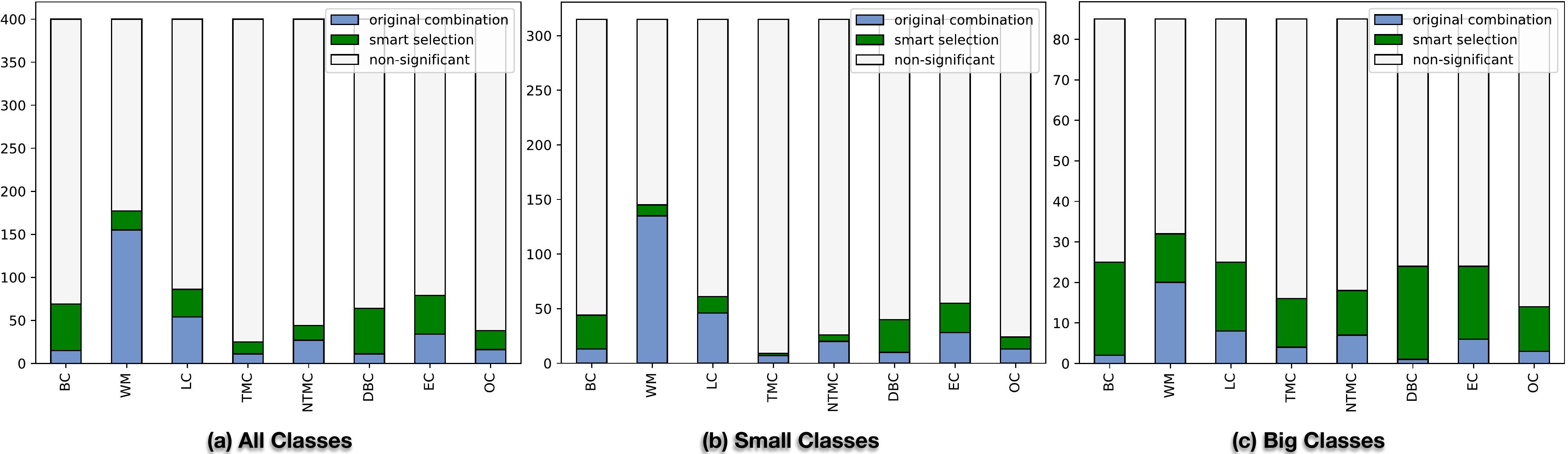}
\caption{Significant case summary of smart selection and the original combination with DynaMOSA}\label{fig:dynamosa_os}
\end{figure*}

\subsection{RQ2: How does smart selection perform with MOSA?}
\noindent\textbf{Motivation.}
In this RQ, we evaluate smart selection with MOSA.

\noindent\textbf{Methodology.}
The methodology is the same as RQ1's.

\noindent\textbf{Result.}
\begin{table}[htbp]
	\centering
	\footnotesize
	\caption{Average coverage results for each approach with MOSA}
	\label{tab:mosa}
\footnotesize{\textbf{(a)} All Classes}\\
\begin{tabular}{l| l| l |l  }
\hline
approach & SS & OC & CC \\ \hline
BC & 57\% & 56\% & \textcolor[RGB]{0,128,28}{58\%} \\ \hline
WM & 60\% & 60\% & 60\% \\ \hline
LC & \textcolor[RGB]{0,128,28}{61\%} & 60\% & \textcolor[RGB]{0,128,28}{61\%} \\ \hline
TMC & \textcolor[RGB]{0,128,28}{84\%} & 83\% & 82\% \\ \hline
\end{tabular}
\begin{tabular}{l| l| l |l  }
\hline
approach & SS & OC & CC \\ \hline
NTMC & \textcolor[RGB]{0,128,28}{71\%} & \textcolor[RGB]{0,128,28}{71\%} & 69\% \\ \hline
DBC & \textcolor[RGB]{0,128,28}{57\%} & 55\% & \textcolor[RGB]{0,128,28}{57\%} \\ \hline
EC & \textcolor[RGB]{0,128,28}{16.95} & 16.15 & 16.41 \\ \hline
OC & 44\% & 44\% & \textcolor[RGB]{0,128,28}{45\%} \\ \hline
\end{tabular}\\
\footnotesize{\textbf{(b)} Small Classes}\\
\begin{tabular}{l| l| l |l  }
\hline
approach & SS & OC & CC \\ \hline
BC & 59\% & 58\% & \textcolor[RGB]{0,128,28}{60\%} \\ \hline
WM & 62\% & 62\% & 62\% \\ \hline
LC & 63\% & 63\% & 63\% \\ \hline
TMC & \textcolor[RGB]{0,128,28}{85\%} & \textcolor[RGB]{0,128,28}{85\%} & 84\% \\ \hline

\end{tabular}
\begin{tabular}{l| l| l |l  }
\hline
approach & SS & OC & CC \\ \hline
NTMC & \textcolor[RGB]{0,128,28}{73\%} & 72\% & 71\% \\ \hline
DBC & 58\% & 58\% & \textcolor[RGB]{0,128,28}{59\%} \\ \hline
EC & \textcolor[RGB]{0,128,28}{13.28} & 13.07 & 12.73 \\ \hline
OC & 45\% & 45\% & \textcolor[RGB]{0,128,28}{46\%} \\ \hline
\end{tabular}\\
\footnotesize{\textbf{(c)} Big Classes}\\
\begin{tabular}{l| l| l |l  }
\hline
approach & SS & OC & CC \\ \hline
BC & 49\% & 46\% & \textcolor[RGB]{0,128,28}{51\%} \\ \hline
WM & \textcolor[RGB]{0,128,28}{52\%} & 49\% & 51\% \\ \hline
LC & \textcolor[RGB]{0,128,28}{54\%} & 52\% & 53\% \\ \hline
TMC & \textcolor[RGB]{0,128,28}{79\%} & 78\% & 77\% \\ \hline
\end{tabular}
\begin{tabular}{l| l| l |l  }
\hline
approach & SS & OC & CC \\ \hline
NTMC & \textcolor[RGB]{0,128,28}{66\%} & 64\% & 64\% \\ \hline
DBC & 50\% & 46\% & \textcolor[RGB]{0,128,28}{51\%} \\ \hline
EC & \textcolor[RGB]{0,128,28}{30.54} & 27.53 & 30.03 \\ \hline
OC & 40\% & 38\% & \textcolor[RGB]{0,128,28}{42\%} \\ \hline
\end{tabular}
\end{table}
\begin{table}[htbp]
    \centering
    \small
    \caption{Average test suite size of each approach with MOSA}
    \begin{tabular}{l|l|l|l}
    \hline
        approach & SS & OC & CC (Average)  \\ \hline
        size (All Classes) & 57.03 & 54.46 & 31.47 \\ \hline
        size (Small Classes) & 38.27 & 38.83 & 19.85 \\ \hline
        size (Big Classes) & 126.56 & 112.38 & 74.53 \\ \hline
    \end{tabular}
    \label{suite_size_mosa}
\end{table}
\noindent$\blacktriangleright$\textbf{All Classes.}

\noindent\textbf{Significant Cases.} Fig. \ref{fig:mosa_os} (a) shows the comparison results of SS and OC on all $400$ Java classes. For each criterion, on average, SS-outperforming classes are $78$ ($19.5\%$). OC-outperforming classes are $42$ ($10.4\%$). No-significant classes are $280$ ($70.1\%$).

\noindent\textbf{Average Coverage.} Table \ref{tab:mosa} (a) shows the average coverage of all classes. SS outperforms OC for five criteria' coverage. Among three approaches, SS reaches five criteria' highest coverage.

\noindent\textbf{Average Suite Size.} The first row of Table \ref{suite_size_mosa} shows the average suite sizes of all classes. Compared to CC, the size of OC increases by $73.1\%$. Compared to OC, the size of SS increases by $4.7\%$.

\noindent$\blacktriangleright$\textbf{Small Classes.}

\noindent\textbf{Significant Cases.} Fig. \ref{fig:mosa_os} (b) shows the comparison results of SS and OC on $315$ small Java classes. For each criterion, on average, SS-outperforming classes are $43$ ($13.8\%$). OC-outperforming classes are $37$ ($11.7\%$). No-significant classes are $235$ ($74.5\%$).

\noindent\textbf{Average Coverage.} Table \ref{tab:mosa} (b) shows the average coverage of small classes. SS outperforms OC for three criteria' coverage. SS reaches three criteria' highest coverage.

\noindent\textbf{Average Suite Size.} The second row of Table \ref{suite_size_mosa} shows the average suite sizes of small classes. Compared to CC, the size of OC increases by $95.6\%$. OC is nearly equal to SS.

\noindent$\blacktriangleright$\textbf{Big Classes.}

\noindent\textbf{Significant Cases.} Fig. \ref{fig:mosa_os} (c) shows the comparison results of SS and OC on $85$ big Java classes. For each criterion, on average, SS-outperforming classes are $35$ ($40.9\%$). OC-outperforming classes are $5$ ($5.6\%$). No-significant classes are $46$ ($53.5\%$).

\noindent\textbf{Average Coverage.} Table \ref{tab:mosa} (c) shows the average coverage of big classes. SS outperforms OC for eight criteria' coverage. SS reaches five criteria' highest coverage.

\noindent\textbf{Average Suite Size.} The third row of Table \ref{suite_size_mosa} shows the average suite sizes of big classes. Compared to CC, the size of OC increases by $50.7\%$. Compared to OC, the size of SS increases by $12.6\%$.

\noindent\textbf{Analysis.}
SS outperforms OC on the big classes like WS. But the advantage of SS is unnoticeable on the small classes. The coverage gap between CC and OC is not significant as the gap in WS. SS nearly bridges this gap. The biggest gap is $5\%$, happening in branch coverage of the big classes. SS narrows this gap by $3\%$. The suite size gap between SS and OC is smaller than on WS, which is consistent with the fact that SS and OC have a smaller coverage gap on MOSA. These facts show the advantage of multi-objective approaches (e.g., MOSA) over single-objective approaches (e.g., WS) \cite{KnowlesSingle, PanichellaMOSA, BrockhoffMulti}. However, the advantage of SS on the big classes indicates that too many objectives also affect the multi-objective algorithms.
\begin{tcolorbox}[title=Answer to RQ2,boxrule=1pt,boxsep=1pt,left=2pt,right=2pt,top=2pt,bottom=2pt]
With MOSA, SS outperforms OC statistically on the big classes. Smart selection has only a slight advantage on the small classes.
\end{tcolorbox}
\begin{figure*}[t]
\centering
\includegraphics[width=1\textwidth]{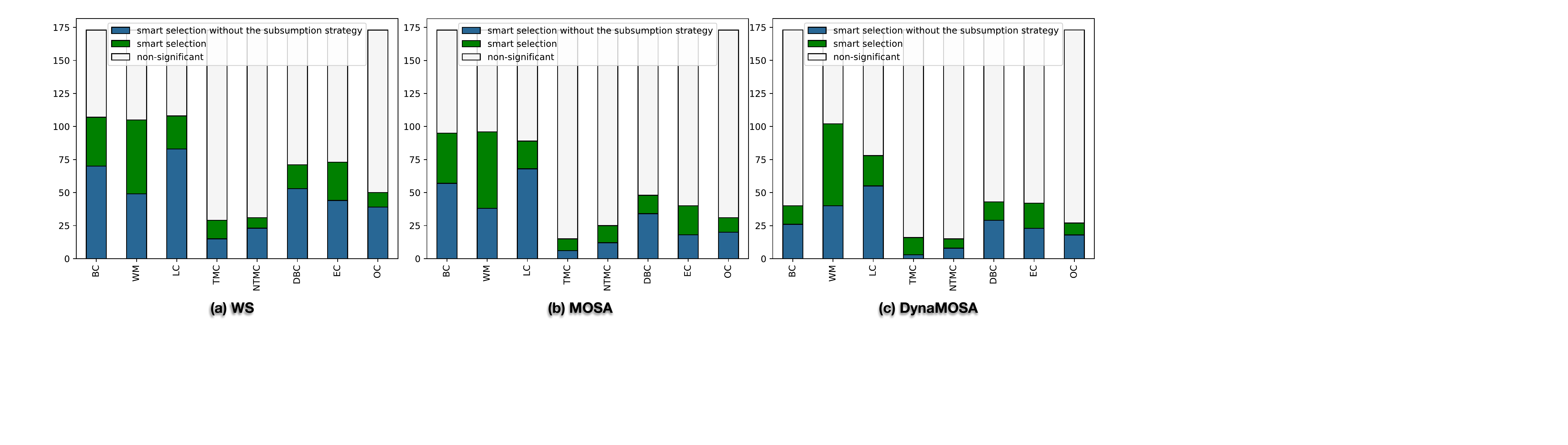}
\caption{Significant case summary of smart selection and smart selection with subsumption strategy}\label{fig:rs}
\end{figure*}
\subsection{RQ3: How does smart selection perform with DynaMOSA?}
\noindent\textbf{Motivation.}
In this RQ, we evaluate smart selection with DynaMOSA.

\noindent\textbf{Methodology.}
The methodology is the same as RQ1's.

\noindent\textbf{Result.}
\begin{table}[htbp]
	\centering
	\footnotesize
	\caption{Average coverage results for each approach with DynaMOSA}
	\label{tab:dynamosa}
	\footnotesize{\textbf{(a)} All Classes}\\
\begin{tabular}{l| l| l |l  }
\hline
approach & SS & OC & CC \\ \hline
BC & 58\% & 58\% & 58\% \\ \hline
WM & 60\% & 61\% & \textcolor[RGB]{0,128,28}{62\%} \\ \hline
LC & 61\% & \textcolor[RGB]{0,128,28}{62\%} & \textcolor[RGB]{0,128,28}{62\%} \\ \hline
TMC & \textcolor[RGB]{0,128,28}{83\%} & \textcolor[RGB]{0,128,28}{83\%} & 81\% \\ \hline
\end{tabular}
\begin{tabular}{l| l| l |l  }
\hline
approach & SS & OC & CC \\ \hline
NTMC & \textcolor[RGB]{0,128,28}{71\%} & \textcolor[RGB]{0,128,28}{71\%} & 70\% \\ \hline
DBC & 57\% & 57\% & \textcolor[RGB]{0,128,28}{58\%} \\ \hline
EC & \textcolor[RGB]{0,128,28}{17.15} & 17.14 & 16.64 \\ \hline
OC & 45\% & 45\% & 45\% \\ \hline
\end{tabular}\\
\footnotesize{\textbf{(b)} Small Classes}\\
\begin{tabular}{l| l| l |l  }
\hline
approach & SS & OC & CC \\ \hline
BC & \textcolor[RGB]{0,128,28}{60\%} & 59\% & \textcolor[RGB]{0,128,28}{60\%} \\ \hline
WM & 63\% & 63\% & \textcolor[RGB]{0,128,28}{64\%} \\ \hline
LC & 63\% & \textcolor[RGB]{0,128,28}{64\%} & \textcolor[RGB]{0,128,28}{64\%} \\ \hline
TMC & 84\% & \textcolor[RGB]{0,128,28}{85\%} & 82\% \\ \hline
\end{tabular}
\begin{tabular}{l| l| l |l  }
\hline
approach & SS & OC & CC \\ \hline
NTMC & 72\% & \textcolor[RGB]{0,128,28}{73\%} & 72\% \\ \hline
DBC & 59\% & 59\% & 59\% \\ \hline
EC & \textcolor[RGB]{0,128,28}{13.42} & \textcolor[RGB]{0,128,28}{13.42} & 12.81 \\ \hline
OC & 46\% & 46\% & 46\% \\ \hline
\end{tabular}\\
\footnotesize{\textbf{(c)} Big Classes}\\
\begin{tabular}{l| l| l |l  }
\hline
approach & SS & OC & CC \\ \hline
BC & 51\% & 51\% & \textcolor[RGB]{0,128,28}{53\%} \\ \hline
WM & 53\% & 52\% & \textcolor[RGB]{0,128,28}{54\%} \\ \hline
LC & 54\% & 54\% & \textcolor[RGB]{0,128,28}{55\%} \\ \hline
TMC & \textcolor[RGB]{0,128,28}{79\%} & \textcolor[RGB]{0,128,28}{79\%} & 78\% \\ \hline
\end{tabular}
\begin{tabular}{l| l| l |l  }
\hline
approach & SS & OC & CC \\ \hline
NTMC & \textcolor[RGB]{0,128,28}{66\%} & \textcolor[RGB]{0,128,28}{66\%} & 65\% \\ \hline
DBC & 51\% & 50\% & \textcolor[RGB]{0,128,28}{53\%} \\ \hline
EC & \textcolor[RGB]{0,128,28}{30.98} & 30.93 & 30.81 \\ \hline
OC & 41\% & 41\% & \textcolor[RGB]{0,128,28}{42\%} \\ \hline
\end{tabular}
\end{table}
\begin{table}[htbp]
    \centering
    \small
    \caption{Average test suite size of each approach with DynaMOSA}
    \begin{tabular}{l|l|l|l}
    \hline
        approach & SS & OC & CC (Average)  \\ \hline
        size (All Classes) & 61.13 & 60.59 & 39.2 \\ \hline
        size (Small Classes) & 38.9 & 39.59 & 23.71 \\ \hline
        size (Big Classes) & 143.51 & 138.44 & 96.58 \\ \hline
    \end{tabular}
    \label{suite_size_dynamosa}
\end{table}
\noindent$\blacktriangleright$\textbf{All Classes.}

\noindent\textbf{Significant Cases.} Fig. \ref{fig:dynamosa_os} (a) shows the comparison results of SS and OC on all $400$ Java classes. For each criterion, on average, SS-outperforming classes are $32$ ($8.1\%$). OC-outperforming classes are $40$ ($10.1\%$). No-significant classes are $328$ ($81.8\%$).

\noindent\textbf{Average Coverage.} Table \ref{tab:dynamosa} (a) shows the average coverage of all classes with three approaches. SS outperforms OC for one criterion's coverage, i.e., exception coverage. Among three approaches, SS reaches the highest coverage for three criteria.

\noindent\textbf{Average Suite Size.} The first row of Table \ref{suite_size_dynamosa} shows the average suite sizes of all classes. Compared to CC, the size of OC increases by $54.7\%$. OC is nearly equal to SS.

\noindent$\blacktriangleright$\textbf{Small Classes.}

\noindent\textbf{Significant Cases.} Fig. \ref{fig:dynamosa_os} (b) shows the comparison results of SS and OC on $315$ small Java classes. For each criterion, on average, SS-outperforming classes are $17$ ($5.2\%$). OC-outperforming classes are $34$ ($10.8\%$). No-significant classes are $265$ ($84\%$).

\noindent\textbf{Average Coverage.} Table \ref{tab:dynamosa} (b) shows the average coverage of small classes. SS outperforms OC for one criterion's coverage (branch coverage). SS reaches two criteria' highest coverage.

\noindent\textbf{Average Suite Size.} The second row of Table \ref{suite_size_dynamosa} shows the average suite sizes of small classes. Compared to CC, the size of OC increases by $66.9\%$. OC is nearly equal to SS.

\noindent$\blacktriangleright$\textbf{Big Classes.}

\noindent\textbf{Significant Cases.} Fig. \ref{fig:dynamosa_os} (c) shows the comparison results of SS and OC on $85$ big Java classes. For each criterion, on average, SS-outperforming classes are $16$ ($18.7\%$). OC-outperforming classes are $6$ ($7.5\%$). No-significant classes are $63$ ($73.8\%$).

\noindent\textbf{Average Coverage.} Table \ref{tab:dynamosa} (c) shows the average coverage of big classes. SS outperforms OC for three criteria' coverage. SS reaches three criteria' highest coverage.

\noindent\textbf{Average Suite Size.} The third row of Table \ref{suite_size_dynamosa} shows the average suite sizes of big classes. Compared to CC, the size of OC increases by $43.3\%$. Compared to OC, the size of SS increases by $3.7\%$.

\noindent\textbf{Analysis.}
SS still outperforms OC on the big classes, but not as obvious as WS and MOSA. In addition, SS is almost the same or slightly worse than OC on the small classes. Furthermore, the coverage gaps among the three approaches are not significant. The gap in the suite size between SS and OC is slight as in the coverage. The reason is that DynaMOSA selects the uncovered goals into the iteration process only when its branch dependencies are covered (see Sec. \ref{subsec:ga}). Hence, the number of optimization objectives is reduced. Therefore, an increase in the goals has a much smaller impact on DynaMOSA's coverage performance than on WS and MOSA.
\begin{tcolorbox}[title=Answer to RQ3,boxrule=1pt,boxsep=1pt,left=2pt,right=2pt,top=2pt,bottom=2pt]
With DynaMOSA, smart selection slightly outperforms the original combination on the big classes. But, the original combination slightly outperforms smart selection on the small classes.
\end{tcolorbox} 
\subsection{RQ4: How does the subsumption strategy affect the performance of smart selection?}\label{subsec:rq4}
\noindent\textbf{Motivation.}
 We select the representative goals from line coverage and weak mutation by the subsumption relationships (see Sec \ref{subsec:rs}). We need to test how it affects the performance of smart selection.

\noindent\textbf{Subjects.}
From $400$ classes, we select those classes that satisfy this condition: The subsumption strategy can find at least one line coverage goal and one mutant. As a result, $173$ classes are selected. 

\noindent\textbf{Configuration.}
We take smart selection without the subsumption strategy (SSWS) as a new approach. Then we keep other configuration settings the same as RQ1-3's.

\noindent\textbf{Methodology.}
To compare SS and SSWS, we follow RQ1's methodology.

\noindent\textbf{Result.}
\begin{table}[htbp]
	\centering
	\footnotesize
	\caption{Average coverage and size results for smart selection and smart selection without the subsumption strategy}
	\label{tab:rq4}
	\footnotesize{\textbf{(a)} WS (Suite Size: SS (45.18), SSWS (48.08))}\\
\begin{tabular}{l| l| l  }
\hline
approach & SS & SSWS \\ \hline
BC & 47\% & \textcolor[RGB]{0,128,28}{48\%} \\ \hline
WM & 52\% & \textcolor[RGB]{0,128,28}{53\%} \\ \hline
LC & 51\% & \textcolor[RGB]{0,128,28}{53\%} \\ \hline
TMC & \textcolor[RGB]{0,128,28}{79\%} & 78\% \\ \hline
\end{tabular}
\begin{tabular}{l| l| l  }
\hline
approach & SS & SSWS \\ \hline
NTMC & 63\% & 63\% \\ \hline
DBC & 46\% & \textcolor[RGB]{0,128,28}{48\%} \\ \hline
EC & 16.43 & \textcolor[RGB]{0,128,28}{17.46} \\ \hline
OC & 38\% & \textcolor[RGB]{0,128,28}{39\%} \\ \hline
\end{tabular}\\
\footnotesize{\textbf{(b)} MOSA (Suite Size: SS (52.08), SSWS (51.66))}\\
\begin{tabular}{l| l| l  }
\hline
approach & SS & SSWS \\ \hline
BC & 49\% & 49\% \\ \hline
WM & 53\% & 53\% \\ \hline
LC & 53\% & \textcolor[RGB]{0,128,28}{54\%} \\ \hline
TMC & \textcolor[RGB]{0,128,28}{79\%} & 78\% \\ \hline
\end{tabular}
\begin{tabular}{l| l| l  }
\hline
approach & SS & SSWS \\ \hline
NTMC & 63\% & 63\% \\ \hline
DBC & 49\% & 49\% \\ \hline
EC & 18.19 & \textcolor[RGB]{0,128,28}{18.32} \\ \hline
OC & 38\% & \textcolor[RGB]{0,128,28}{39\%} \\ \hline
\end{tabular}\\
\footnotesize{\textbf{(c)} DynaMOSA (Suite Size: SS (57.1), SSWS (54.79))}\\
\begin{tabular}{l| l| l  }
\hline
approach & SS & SSWS \\ \hline
BC & 50\% & 50\% \\ \hline
WM & 54\% & 54\% \\ \hline
LC & 53\% & \textcolor[RGB]{0,128,28}{54\%} \\ \hline
TMC & \textcolor[RGB]{0,128,28}{79\%} & 78\% \\ \hline
\end{tabular}
\begin{tabular}{l| l| l  }
\hline
approach & SS & SSWS \\ \hline
NTMC & 63\% & 63\% \\ \hline
DBC & 49\% & 49\% \\ \hline
EC & 18.26 & \textcolor[RGB]{0,128,28}{18.68} \\ \hline
OC & 39\% & 39\% \\ \hline
\end{tabular}
\end{table}
\noindent$\blacktriangleright$\textbf{WS.}

\noindent\textbf{Significant Cases.} Fig. \ref{fig:rs} (a) shows the comparison results of SS and SSWS on $173$ classes with WS. For each criterion, on average, SS-outperforming classes are $25$ ($14.5\%$).  SSWS-outperforming classes are $47$ ($27.2\%$). No-significant classes are $101$ ($58.3\%$).

\noindent\textbf{Average Coverage.} Table \ref{tab:rq4} (a) shows the average coverage for WS. SS outperforms SSWS for one criterion's coverage (top-level method coverage). SSWS outperforms SS for six criteria.

\noindent$\blacktriangleright$\textbf{MOSA.}

\noindent\textbf{Significant Cases.} Fig. \ref{fig:rs} (b) shows the comparison results of SS and SSWS on $173$ classes with MOSA. For each criterion, on average, SS-outperforming classes are $23$ ($13.3\%$). SSWS-outperforming classes are $32$ ($18.5\%$). No-significant classes are $118$ ($68.2\%$). 

\noindent\textbf{Average Coverage.} Table \ref{tab:rq4} (b) shows the average coverage for MOSA. SS outperforms SSWS for one criterion's coverage (top-level method coverage). SSWS outperforms SS for three criteria.

\noindent$\blacktriangleright$\textbf{DynaMOSA.}

\noindent\textbf{Significant Cases.} Fig. \ref{fig:rs} (c) shows the comparison results of SS and SSWS on $173$ Java classes with DynaMOSA. On average, SS-outperforming classes are $20$ ($11.6\%$). SSWS-outperforming classes are $25$ ($14.5\%$). No-significant classes are $128$ ($73.9\%$).

\noindent\textbf{Average Coverage.} Table \ref{tab:rq4} (c) shows the average coverage for DynaMOSA. SS outperforms SSWS for one criterion's coverage (top-level method coverage). SSWS outperforms SS for two criteria.

\noindent\textbf{Analysis.}
SSWS outperforms slightly SS for most criteria on WS. It confirms that an increase in the objectives has a much bigger impact on WS than on MOSA and DynaMOSA. Furthermore, the results are different on line coverage and weak mutation for which SS adds subsets. For three algorithms, SSWS is better in line coverage in terms of the outperforming classes and average coverage. Contrarily, SS is better in weak mutation in terms of the outperforming classes. It indicates that the coverage correlation between (direct) branch coverage and line coverage is stronger than the one between (direct) branch coverage and weak mutation. As for the suite size, Table \ref{tab:rq4} shows that SS and SSWS are similar in all three algorithms.

Unexpectedly, SS outperforms SSWS on top-level method coverage. We analyze some classes qualitatively. For example, there is a public method named \textit{compare} in an inner class of the class \textit{org.apache.hadoop.mapred} \cite{HadoopCompareMethod}. The results show that $88$ out of $90$ test suites generated by SS cover this top-level method goal, while only $1$ out of $90$ test suites generated by SSWS covers this goal. We find that this method contains $8$ lines, $2$ branches, and $3$ output goals. EvoSuite skips the branches and output goals in the inner class (lines, methods, and mutants are kept). This class has $350$ branches, $48$ methods, and $10$ output goals. SS selects $14$ lines and $216$ mutants for this class ($0$ lines and $10$ mutants for this method). As a result, if a test directly invokes this method, under SS, at most $(1+10)$ out of $638$ ($2\%$) goals are closer to being covered; under SSWS, the number is $1$ out of $408$ ($0.2\%$). It explains why all algorithms with SSWS tend to ignore this method goal since the gain is tiny. We find that this scenario is common in big classes containing short and branch-less methods.
\begin{tcolorbox}[title=Answer to RQ4,boxrule=1pt,boxsep=1pt,left=2pt,right=2pt,top=2pt,bottom=2pt]
Smart selection without the subsumption strategy outperforms slightly smart selection in most criteria on WS (except for WM and TMC). Smart selection outperforms slightly smart selection without the subsumption strategy in WM and TMC on three algorithms.
\end{tcolorbox}
\section{Discussion}

\subsection{Parameter Tuning} \label{subsec:pt}
Smart selection introduces a new parameter: \textit{lineThreshold} (see Sec. \ref{subsec:rs}). In handling line coverage, smart selection skips those basic blocks (BBs) with lines less than \textit{lineThreshold}. The larger the value of this parameter, the more BBs we skip. Without considering the dead code, (direct) branch coverage fails to capture the following lines only when a certain line in a basic block exits abnormally. Previous work \cite{Rojas2015CombiningMC} shows that, on average, when $78\%$ of branches are covered, test suites can only find $1.75$ exceptions. It indicates that (direct) branch coverage can capture most properties of line coverage. Therefore, to minimize the impacts of line coverage goals on SBST, we prefer a larger \textit{lineThreshold}. After statistics on the benchmarks used in DynaMOSA \cite{PanichellaDynaMOSA}, we find that $50\%$ of the BBs have less than $8$ lines. Therefore, we set \textit{lineThreshold} to $8$.

\subsection{Threats to Validity}
The threat to external validity comes from the experimental subjects. We choose 158 Java classes from the benchmark of DynaMOSA \cite{PanichellaDynaMOSA}. \cite{PanichellaDynaMOSA} was published in 2018. Many classes have already become obsolete. Some projects even are no longer maintained \cite{LinGraph}. To reduce this risk, we choose 242 classes at random from Hadoop \cite{hadoop}, thereby increasing the diversity of the dataset. The threat to internal validity comes from the randomness of the genetic algorithms. To reduce the risk, we repeat each approach 30 times for every class.

\section{Related Work}
In this section, we introduce related studies on (1) SBST and (2) coverage criteria combination in SBST.

\noindent\textbf{SBST.} 
SBST formulates test cases generation as an optimization problem. Miller et al. \cite{miller1976automatic} proposed the first SBST technique to generate test data for functions with inputs of float type. SBST techniques have been widely used in various objects under test \cite{FraserEvoSuite, arcuri2018evomaster, castelein2018search, gambi2019automatically, wang2021ML,Tang:2020,dong2020time,martin2021restest, haq2021automatic}, and types of software testing \cite{li2007regression, silva2017systematic, walcott2006timeaware}. Most researchers focus on (1) search algorithms: Tonella \cite{TonellaEvo2004} proposed to iterate to generate one test case for each branch. Fraser et al. \cite{FraserWhole} proposed to generate a test suite for all branches. Panichella et al. \cite{PanichellaMOSA, PanichellaDynaMOSA} introduced many-objective optimization algorithms. Grano et al. \cite{grano2019testing} proposed a variant of DynaMOSA \cite{PanichellaDynaMOSA} to reduce the computation costs; (2) fitness gradients recovery: Lin et al. \cite{lin2020recovering} proposed an approach to address the inter-procedural flag problem. Lin et al. \cite{LinGraph} proposed a test seed synthesis approach to create complex test inputs. Arcuri et al. \cite{arcuri2021enhancing} integrated testability transformations into API tests. Braione et al. \cite{braione2017combining} combined symbolic execution and SBST for programs with complex inputs; (3) readability of generated tests: Daka et al. \cite{daka2017generating} proposed to assign names for tests by summarizing covered coverage goals. Roy et al. \cite{roy2020deeptc} introduced deep learning approaches to generate test names; (4) fitness function design: Xu et al. \cite{xu2017adaptive} proposed an adaptive fitness function for improving SBST. Rojas et al. \cite{Rojas2015CombiningMC} proposed to combine multiple criteria to satisfy users' requirements.

\noindent\textbf{Coverage Criteria Combination in SBST.}
Rojas et al. \cite{Rojas2015CombiningMC} proposed to combine multiple criteria to guide SBST. Gregory Gay \cite{GayCombine} experimented with different combinations of coverage criteria. His experiment compares the effectiveness of multi-criteria suites in detecting complex, real-world faults. Omur et al. \cite{SAHIN2021806} introduced the Artificial Bee Colony algorithm as a substitute for the genetic algorithms used in WS \cite{FraserWhole}. Our work aims to increase the coverage decrease caused by combing multiple criteria \cite{Rojas2015CombiningMC} and is orthogonal to the latter two studies \cite{GayCombine, SAHIN2021806}.
\section{Conclusion}
We propose smart selection to address the coverage decrease caused by combining multiple criteria in SBST. We compare smart selection with the original combination on $400$ Java classes. The experiment results confirm that with WS and MOSA, smart selection outperforms the original combination, especially for the Java classes with no less than $200$ branches. But with DynaMOSA, the differences between smart selection and the original combination are slight.
\section{Acknowledgements}
Z. Zhou and Y. Tang are partially sponsored by Shanghai Pujiang Program (No. 21PJ1410700). Y. Zhou is partially sponsored by National Natural Science Foundation of China (No. 62172205). C. Fang and Z. Chen are partially sponsored by Science, Technology and Innovation Commission of Shenzhen Municipality \newline (CJGJZD20200617103001003)

\bibliographystyle{ACM-Reference-Format}
\bibliography{ref.bib}

\end{document}